\documentclass[aps,prb,reprint,superscriptaddress,longbibliography]{revtex4-1}
\usepackage{amssymb}
\usepackage[utf8]{inputenc}
\usepackage{graphicx}
\usepackage{bm,color,subfigure,amsmath}

\begin{document}

\title{Controllable vortex shedding from dissipative exchange flows in ferromagnetic channels}

\author{Ezio~Iacocca}
\email{ezio.iacocca@northumbria.ac.uk}
\affiliation{Department of Mathematics, Physics, and Electrical Engineering, Northumbria University, Newcastle upon Tyne, NE1 8ST, UK}

\begin{abstract}
Ferromagnetic channels subject to spin injection at one extremum sustain long-range coherent textures that carry spin currents known as dissipative exchange flows (DEFs). In the weak injection regime, spin currents carried by DEFs decay algebraically and extend through the length of the channel, a regime known as spin superfluidity. Similar to fluids, these structures are prone to phase-slips that manifest as vortex-antivortex pairs. Here, we numerically study vortex shedding from DEFs excited in a magnetic nanowire with a physical obstacle. Using micromagnetic simulations, we find regimes of laminar flow and vortex shedding as a function of obstacle position tunable by the and spin injection sign and magnitude. Vortex-antivortex pairs translate forward (VF regime) or backward (VB regime) with respect to the detector's extremum, resulting in well-defined spectral features. Qualitatively similar results are obtained when temperature, anisotropy, and weak non-local dipole fields are included in the simulations. These results provide clear features associated with DEFs that may be detected experimentally in devices with nominally identical boundary conditions. Furthermore, our results suggest that obstacles can be considered as DEF control gates, opening an avenue to manipulate DEFs via physical defects.
\end{abstract}

\maketitle

\section{Introduction}

Spin transport through magnetic materials has attracted considerable interest due to the possibility of manipulating information encoded by angular momentum. Spin waves carry angular momentum and can be electrically excited by, e.g., spin-transfer torque~\cite{Ralph2008} or spin-Hall effect~\cite{Hoffmann2013}. Spin waves typically propagate for several hundred nanometers, a length that is limited by the spin waves' frequency and the material's magnetic damping~\cite{Madami2011}. By utilizing materials with low damping, spin waves have been detected beyond a micrometer, for example, in ferromagnetic permalloy (Py)~\cite{Madami2011,Cornelissen2015}, ferrimagnetic yttrium iron garnet (YIG)~\cite{Wesenberg2017,Liu2018}, and antiferromagnetic haematite~\cite{Lebrun2018}.

A superior means to transport angular momentum through magnetic materials was theoretically envisioned as a spin supercurrent~\cite{Konig2001}, a nonlinear magnetic texture composed of a continuous rotation of the magnetization about the normal-to-plane axis. Because the magnetization texture is homochiral, its order parameter is analogous to that of a supercurrent. This feature sparked intense theoretical research that predicted current-induced spin superfluids or dissipative exchange flows (DEFs) in magnetic nanowires~\cite{Sonin2010,Takei2014,Takei2014b,Chen2014,Skarsvaag2015,Sonin2017,Iacocca2017d,Schneider2018,Sonin2019,Iacocca2019b,Evers2020}. Collectively, such textures can be analytically studied within the context of spin hydrodynamics~\cite{Iacocca2017,Iacocca2017b,Iacocca2019_Rev} beyond the traditional long-wave approximation~\cite{Halperin1969}.

DEFs are promising for long-distance spin transport because its spatial profile is set by boundary conditions~\cite{Iacocca2019b}. This means that DEFs can theoretically extend through arbitrary distances, well above a micrometer. Once a DEF is stabilized in a nanowire, the texture can only unwind through phase slips~\cite{Kim2016,Kim2016b,Iacocca2017b}. The continual precessional motion of the magnetization in the GHZ range effectively translates the DEF along the nanowire~\cite{Iacocca2019b}. The precessional frequency can be read-out at the nanowire's detector extremum as spin pumping~\cite{Takei2014}.

Experimentally, stabilizing a DEF has been a difficult task. There are two pieces of experimental evidence to date. Yuan et al.~[\onlinecite{Yuan2018}] studied spin injection in non-local antiferromagnetic Cr$_2$O$_3$ devices. By varying the distance between the injector and detector Pt strips, the non-local resistance trend was fitted to an algebraic decay. Stepanov et al.~[\onlinecite{Stepanov2018}] studied a more exotic graphene quantum Hall antiferromagnet to realize an almost ideal spin injection in a non-local device. They detected a non-local signal at over 5~$\mu$m, attributing such a strong transport to a spin superfluid. While these studies provide evidence that long-distance spin transport cannot be entirely attributed to spin waves, conclusive evidence supporting the existence of DEFs or spin superfluids remains elusive.

The main feature of linear DEFs is their algebraically decaying profile as opposed to the exponential decay of spin waves. However, it must be recognized that this is a consequence of the boundary conditions that ultimately determine the DEF's profile and frequency~\cite{Takei2014,Iacocca2019b,Evers2020}. The practical consequence is that measurements relying on varying the detector's position influence the DEF's profile and, therefore, the detected signal is prone to a variety of factors~\cite{Lebrun2018}. Consequently, it is desirable to find evidence of DEFs in devices where the distance between the injector and detector is nominally identical. In this paper, we propose a method to obtain qualitative evidence for DEFs in devices with identical boundary conditions. We are inspired by a previous theoretical work~\cite{Iacocca2017b} where a sizable defect, such as a physical obstacle or a local perpendicular magnetic field, induced vortex-antivortex pair production from a ``uniform hydrodynamic state'' or spin supercurrent~\cite{Konig2001}.

\begin{figure}[t]
\centering \includegraphics[trim={0in 0.1in 0in 0.05in}, clip, width=2.6in]{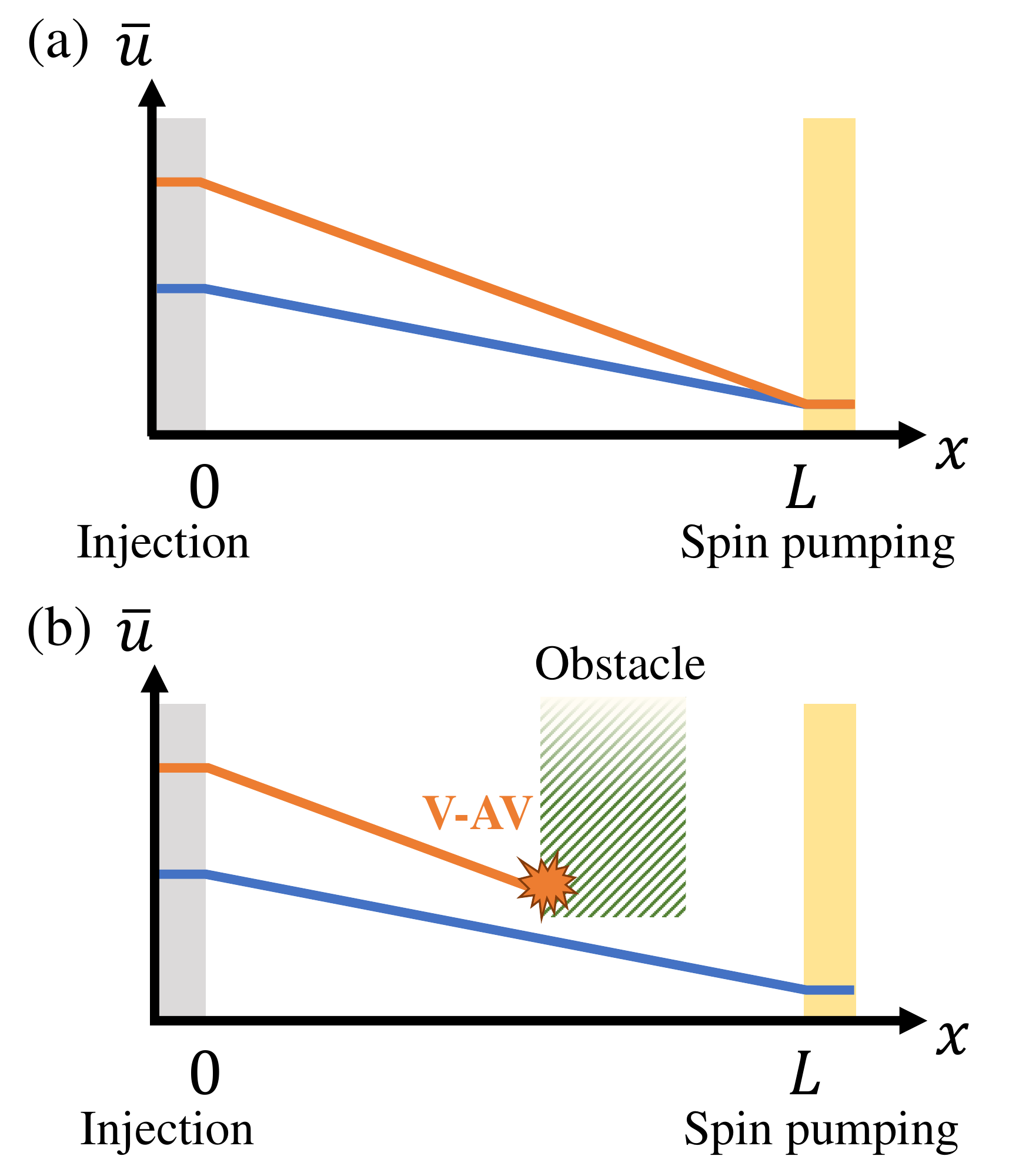}
\caption{ \label{fig1} Schematic of linear DEF solutions in a 1D channel subject to spin injection at $x=0$ (gray region) and spin pumping at $x=L$ (yellow region). (a) In a perfect channel, the DEF's wavevector $\bar{u}$ decays algebraically. (b) When an obstacle is introduced in the channel, the DEF's wavevector is interrupted. In a nanowire, this event is associated with the development of locally supersonic flow that gives rise to vortex-antivortex (V-AV) pair production. }
\end{figure}

In the linear DEF or spin superfluid approximation, a spin injection at $x=0$ gives rise to a structure with a wavevector $\bar{u}$ that then decays algebraically with distance~\cite{Sonin2010,Takei2014,Iacocca2019b}, as schematically shown in Fig.~\ref{fig1}(a) for a 1D channel. A spin detector at position $L$ would then pick-up a signal proportional to the DEF's frequency. A physical defect or obstacle can be thought of as a wavevector-dependent barrier for a DEF, as schematically illustrated in Fig.~\ref{fig1}(b). Above a critical spin injection, the obstacle will induce locally supersonic flow that results in vortex-antivortex (V-AV) pair production, destroying the long-range coherence of the DEF and leading to a drop in the detected signal.

In this paper, we study the characteristic features of DEFs established in nanowires with physical obstacles by micromagnetic simulations. We show that an obstacle can serve as a control gate that destroys the DEF's long-range coherence over a critical injection current that depends on the distance between the obstacle and the injection site. Because the boundary conditions are kept identical, this method can be used to find experimental evidence for DEFs in nanowires where only the position of the obstacle is varied. The results presented here pave the way for the manipulation of long-distance spin transport by obstacles of different shapes, sizes, and arrangements.
\begin{figure}[t]
\centering \includegraphics[trim={0.2in 0in .5in 0.2in}, clip, width=3.4in]{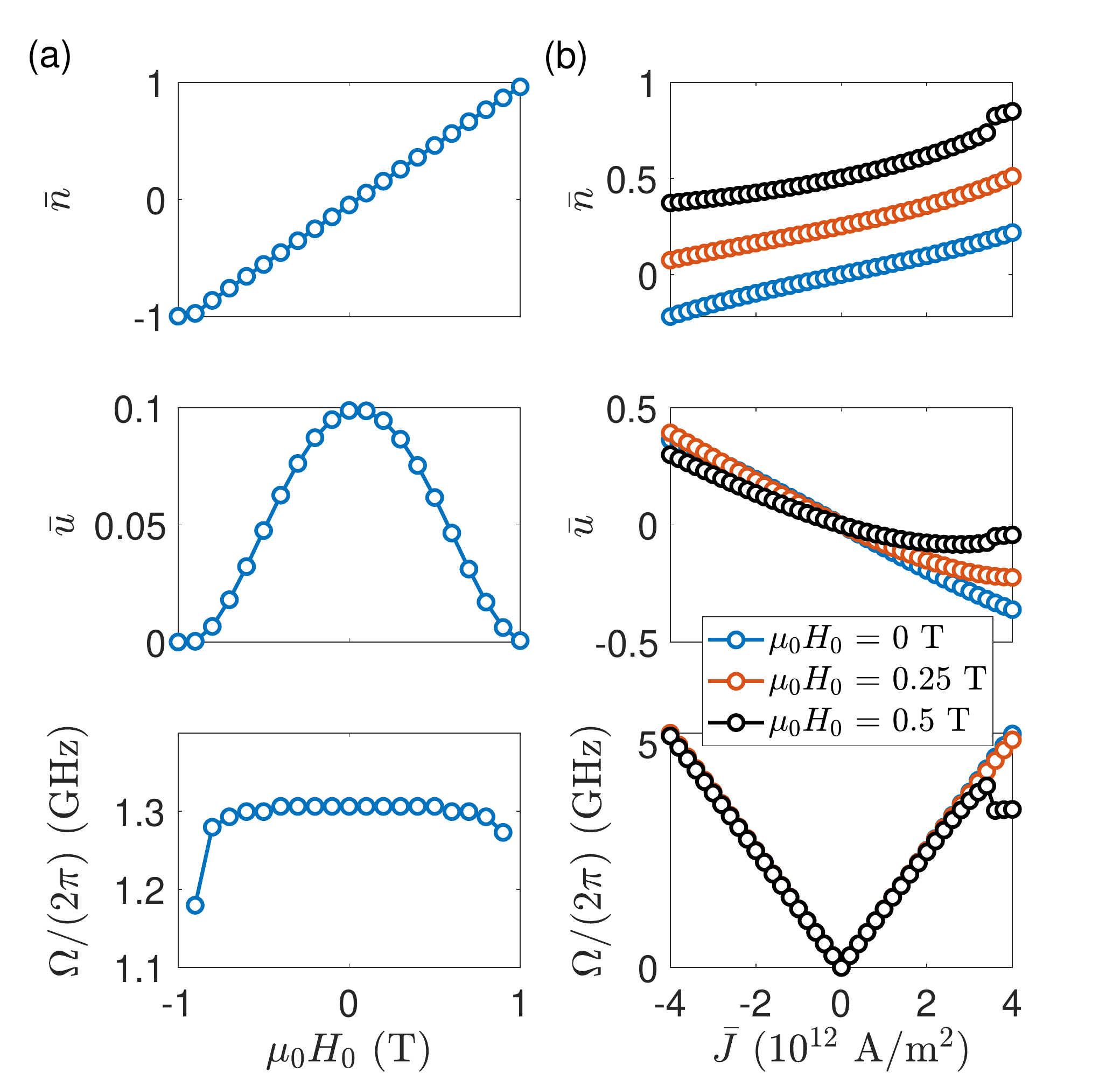}
\caption{ \label{fig3} (a) Field-dependent DEF parameters excited by a current of $\bar{J}=-10^{12}$~A/m$^2$. When the field approaches 1~T, the magnetization saturates and the DEF precessional frequency is not defined. (b) Current-dependent DEF parameters for no external field (blue), $\mu_0H_0=0.25$~T (red), and $\mu_0H_0=0.5$~T (black). A nonlinear dependence of the DEF parameters is observed when the field is non-zero. }
\end{figure}

\section{Current-induced DEF}

The common theoretical setting for stabilizing and detecting DEFs requires a spin injector at position $x=0$ (left) and a spin detector at position $x=L$ (right), as schematically shown in Fig.~\ref{fig1}. Spin injection excites and sustains the DEF while the magnetization precession is detected at the right extremum. We simulate this configuration as a nanowire of length $1024$~nm, width $256$~nm, and thickness $5$~nm by micromagnetic modeling using the MuMax3 package~\cite{Vansteenkiste2014}. We utilize magnetic parameters of permalloy (Py), namely, saturation magnetization $M_s=790$~kA/m, exchange constant $A=10$~pJ/m, and Gilbert damping coefficient $\alpha=0.01$. These material parameters establish an exchange length $\lambda_\mathrm{ex}\approx5$~nm. Unless otherwise specified, the non-local dipole field is disabled. Easy-plane anisotropy is enforced as a negative uniaxial anisotropy field of magnitude $M_s$ along the z-axis (normal-to-plane).

Following the procedure of Ref.~\onlinecite{Iacocca2019b}, we simulate spin injection as spin-transfer torque acting on a $5$-nm-wide region from the left extremum. For simplicity, we set a symmetric ($\lambda=1$) Slonczewski torque with polarization $P=1$ and a current spin-polarized along the $z$-axis. This current exerts a torque that tilts the magnetization vector out of the plane, allowing the internal field to drive its precession. By exchange interaction, the delayed precession of neighboring spins gives rise to a DEF. The DEF's translation is energetically degenerate in either hemisphere of the magnetization's unit sphere. To lift this degeneracy, we apply an external field, $H_0$, along the $z$-axis. We also simulate spin pumping as $5$-nm-wide region with increased damping in the opposite extremum of the nanowire~\cite{Schneider2018}. This implies that the effective 1D channel has a length of $1014$~nm. The dimensionless length $1014$~nm$/\lambda_\mathrm{ex}\approx 200$ is sufficient to ensure a linear DEF solution for low to moderate spin injection strengths~\cite{Iacocca2019b}.

To characterize the DEF, we utilize a dispersive hydrodynamic formulation of magnetization dynamics~\cite{Iacocca2017}. In such a formulation, the magnetization vector $\mathbf{m}=(m_x,m_y,m_z)$ is expressed as a spin density $n=m_z$ and a fluid velocity $u=-\partial_x\Phi=-\partial_x\arctan{(m_y/m_x)}$. DEFs can be parametrized by three quantities~\cite{Iacocca2019b}: the spin density at the injection site $\bar{n}$, the fluid flow injection $\bar{u}$ that is proportional to the injection current, and the uniform precessional frequency $\Omega=\gamma \mu_0M_s\partial_t\Phi$, where $\gamma$ is the gyromagnetic ratio and $\mu_0$ is the vacuum permeability.

We numerically determine the DEF parameters as a function of the simulated spin-polarized charge current density and applied field in Fig.~\ref{fig3}. Panel (a) shows $\bar{n}$, $\bar{u}$, and $\Omega$ as a function of field for a current density of $\bar{J}=-10^{12}$~A/m$^2$. The linear dependence of $\bar{n}$ on $H_0$ is consistent with the field magnetizing the nanowire. The sinusoidal dependence on $\bar{u}$ follows from the symmetric spin-transfer torque angular dependence between the magnetization vector and the polarizer~\cite{Ralph2008}. We note that $\bar{u}$ is maximal at no applied field, reaching a magnitude of $0.1$, within the linear DEF regime~\cite{Iacocca2019b}. Furthermore, we note that the negative current induces a positive $\bar{u}$ in our geometry, were positive $\bar{u}$ is defined as the fluid flow injection directed from the injection site (left) to the spin pumping site (right). In other words, $\bar{u}>0$ indicates flow downstream and $\bar{u}<0$ indicates flow upstream. The DEF's frequency is determined by computing the Fourier transform of the spatially averaged in-plane magnetization over the region subject to spin pumping. We find that the DEF's frequency is approximately constant with the field at $1.3$~GHz. As the field saturates the magnetization out of the plane, $\bar{n}=\pm1$, a DEF solution becomes unavailable so that the frequency decreases until it can no longer be computed.
\begin{figure*}[t]
\centering \includegraphics[trim={1in 0.8in .5in 0.5in}, clip, width=7in]{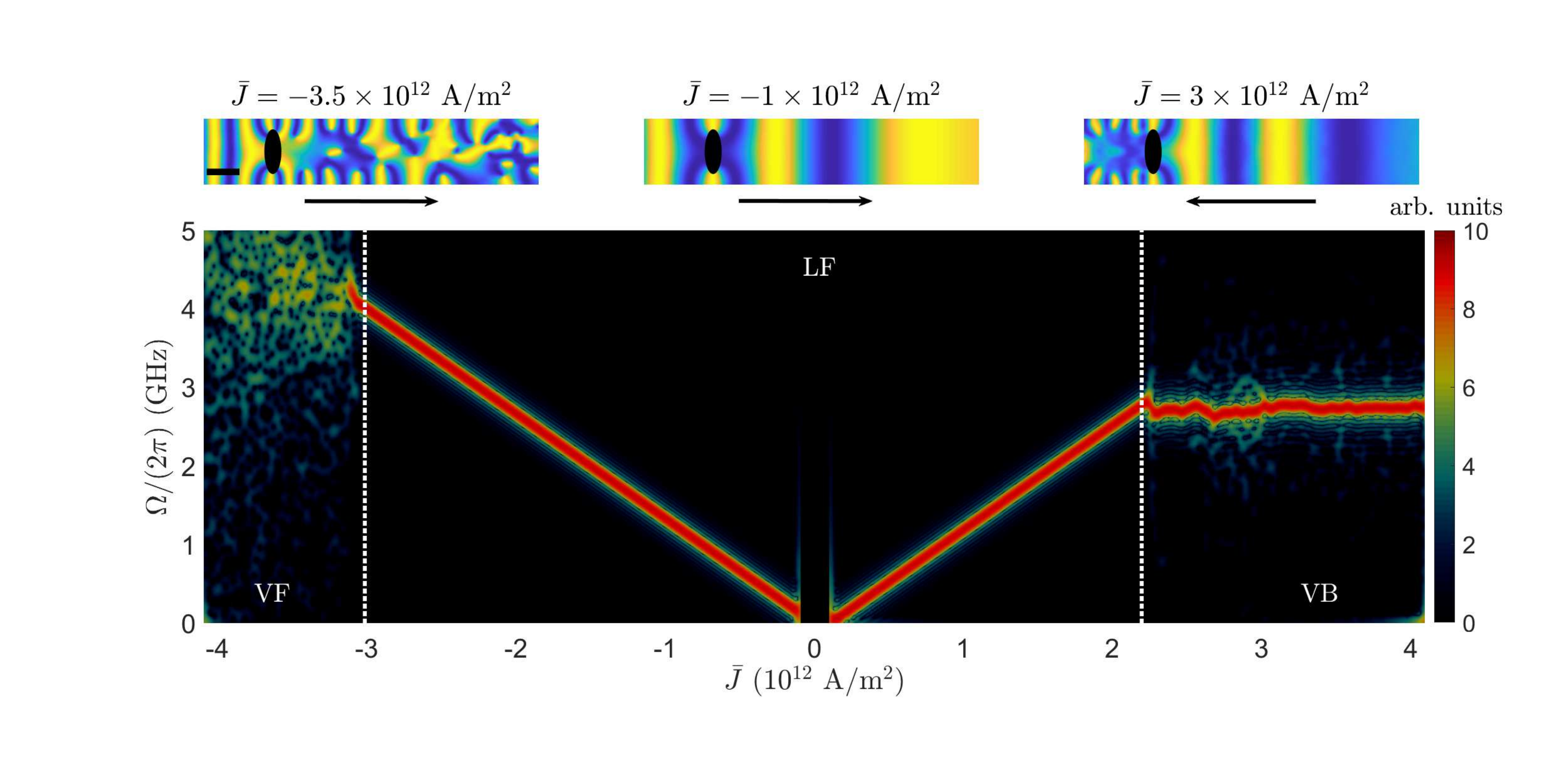}
\caption{ \label{fig4} DEF frequency as a function of current in a nanowire with an obstacle placed 207~nm from the injection site and under a field $\mu_0H_0=0.3$~T. The colormap shown in logarithmic scale is obtained by sweeping the current and performing a time-frequency analysis employing a smoothed pseudo-Wigner-Ville distribution. Three dynamical regimes are identified and separated by vertical white dashed lines. The plot is dominated by laminar flow (LF regime), indicating that the DEF is not altered by the obstacle, shown in the top-middle panel. For large negative currents, V-AV pairs shed towards the detector, or forward (VF regime), as shown in the top-left panel. The spectrum is characterized by noisy, broad features stemming from vortex and antivortex interaction and annihilation events. For large positive currents, V-AV pairs shed against the detector, or backward (VB regime), as shown by the top-right panel. The spectrum is approximately constant indicating a filter-like effect produced by the obstacle. In the top panels, the black arrows indicate the fluid flow direction, showing that V-AV pairs in all cases translate in the direction of the fluid flow}.
\end{figure*}

In panel (b) the DEF parameters are computed as a function of current in the range $-4\times10^{12}$~A/m$^2$ to $4\times10^{12}$~A/m$^2$ for fields of $0$~T, $0.25$~T, and $0.5$~T. Both $\bar{n}$ and $\bar{u}$ are asymmetric with respect to the current's sign and develop nonlinearities as both the current and field increase. The asymmetry and nonlinearity are less pronounced in the frequency, whose linear dependence with current indicates that the simulated parameters are approximately within the linear DEF or spin superfluid regime. In addition, $|\bar{u}|$ does not exceed 0.5 for any field condition, ensuring that regimes of modulational instability due to wavelengths on the order of the exchange length~\cite{Iacocca2017} or contact-soliton DEFs arising from supersonic flow at the injection site~\cite{Iacocca2019b} are avoided in these simulations.

\section{Nanowire with obstacle}

In this section, we investigate the effect of an obstacle on the DEF dynamics. We first consider an elliptical obstacle modeled as a void in the nanowire of dimensions $150$~nm$\times~50$~nm. The long axis is directed along the $y$ direction. These particular dimensions were chosen to enhance the deflected fluid velocity. We place the obstacle at horizontal distances relative to the geometrical center of the nanowire that were set to $100$~nm, $0$~nm, $-100$~nm, $-200$~nm, and $-300$~nm. These correspond to distances $d$ of 607~nm, 507~nm, 407~nm, 307~nm, and 207~nm relative to the injection site, respectively.

We numerically investigate the dynamics as a function of current by setting $d$ and an external field between $0.1$~T and $0.5$~T. For each case, the equilibrium magnetization state is obtained before exciting dynamics with the current. We simulate current densities with both positive and negative polarities, in the range $|\bar{J}|=10^{12}$~A/m$^2$ to $|\bar{J}|=4\times10^{12}$~A/m$^2$. For each sign of the current density, the simulation runs for 500~ns. We then analyze the spatially averaged in-plane magnetization in the region subject to spin pumping with a time-frequency procedure. We employ the smoothed pseudo-Wigner-Ville (SPWV) distribution which minimizes the numerical artifacts of a Wigner-Ville distribution by applying window functions in time and frequency. Here, we implement a Hann frequency window of 0.1~GHz and a Gaussian temporal window of 10~ns, similar to the SPWV implementation used in Ref.~[\onlinecite{Dumas2013}].

The main panel in Fig.~\ref{fig4} shows a color plot of the time-dependent frequency computed with the SPWV distribution described above for the case of $d=207$~nm and $\mu_0H_0=0.3$~T. Three distinct dynamical regimes are found. In the range $\bar{J}=-4\times10^{12}$~A/m$^2$ to $\bar{J}\approx-3\times10^{12}$~A/m$^2$, the spectrum is noisy. This indicates that the magnetization is not homogeneously precessing at the spin pumping extremum. The top left panel shows a snapshot of the $x$ magnetization component for $\bar{J}=-3.5\times10^{12}$~A/m$^2$. The magnetization dynamics are dominated by vortex shedding carried by the fluid flow (black arrow) towards the detector, or forward (VF), that ultimately leads to vortex-antivortex interactions and annihilation events characterized by spin-wave bursts. In the range $\bar{J}\approx-3\times10^{12}$~A/m$^2$ to $\bar{J}\approx2.2\times10^{12}$~A/m$^2$, the spectrum is clean and the frequency is linearly proportional to the current density, in agreement with ideal DEFs, c.f. Fig.~\ref{fig3}(b). As confirmed by the top-center snapshot at $\bar{J}=-10^{12}$~A/m$^2$, this regime is characterized by ``laminar flow'' (LF) around the obstacle that ensures a coherent magnetization precession at the spin pumping site. In the range $\bar{J}\approx2.2\times10^{12}$~A/m$^2$ to $\bar{J}=4\times10^{12}$~A/m$^2$, the frequency is essentially constant. As shown in the top-right snapshot at $\bar{J}=3\times10^{12}$~A/m$^2$, this is a case of vortex shedding where the fluid flow carries them against the detector, or backward (VB). This situation approximately maintains the DEF's coherence at the spin pumping site.

Another distinctive feature between the VF and VB regimes is their spectral characteristics see, e.g., Fig.~\ref{fig4}. In the VF regime, the spectrum is noisy, stemming from V-AV dynamics and annihilation events downstream that destroy the long-range coherence of a DEF. However, the VB regime exhibits a relatively flat spectrum. Because vortices are shed back to the injection site, dynamics and annihilation events result in an incoherent spin injection. One can understand such injection as a random process, where a band of wavelengths is excited. The obstacle then serves as a low-pass filter where only the wavelengths below the VB regime transition travel towards the detector in the form of a DEF. Because the DEF's frequency is proportional to $\bar{u}$, the largest frequency is preferentially detected, modulated by slower frequencies as they approach the spin pumping site. This is precisely what is observed in Fig.~\ref{fig4}, where the VB spectrum feature is approximately flat and exhibits incoherent sidebands in time, implying that the signal is randomly modulated.

The spectral features of each regime can be directly accessible by experiments. Detection methods based on inverse spin-Hall effect~\cite{Yuan2018} rely on the precessing magnetization to pump a spin current $Q_s$ into a spin reservoir~\cite{Tserkovnyak2002b}. In other words, $Q_s\propto\Omega$. This simple proportionality can be generalized for a spectral distribution. For example, a Gaussian lineshape with amplitude $C$ and standard deviation $\sigma$ would lead to a distribution of spin currents with a mean value $\langle Q_s\rangle\propto C\Omega$ and a standard deviation proportional to $\sigma$. This situation practically implies that the detected direct current by inverse spin-Hall effect in the VF regime will be significantly lower than that detected in the LF regime. Furthermore, the transition should be evidenced by a sudden drop in the detected current. In the case of the VB regime, the detected current would plateau as a function of spin injection.
\begin{figure}[t]
\centering \includegraphics[trim={.2in 0in .2in 0.2in}, clip, width=3.4in]{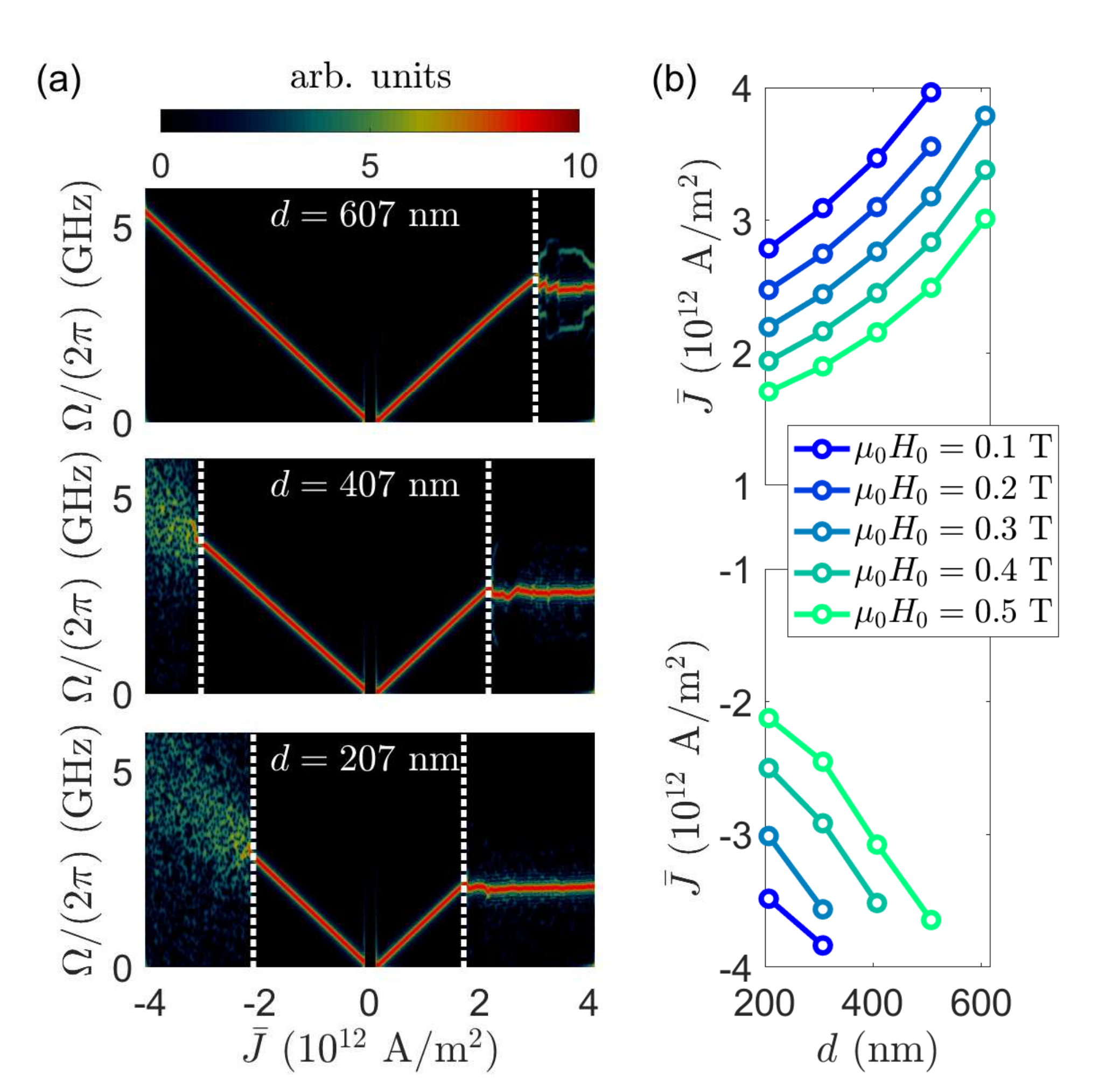}
\caption{ \label{fig5} (a) DEF frequency as a function of current when $\mu_0H_0=0.5$~T shown in logarithmic scale. The obstacle is placed at $607$~nm, $407$~nm, and $207$~nm for the plots shown from top to bottom. The VF and VB regimes' range increase as the obstacle is placed closer to the injection site, where the fluid velocity is larger. (b) Transition currents between laminar flow and VB regime (top panel) and VF regime (bottom panel) as a function of obstacle distance from the injection site and applied field. The perpendicular field eases vortex shedding. The VF regime is not observed for $\mu_0H_0=0.2$~T.}
\end{figure}

The transition between LF and either the VF or VB regimes is tunable by both the obstacle position and the external field. In Fig.~\ref{fig5}(a), we show the time-frequency plot for the spatially averaged in-plane magnetization dynamics at the spin pumping site for $\mu_0H_0=0.5$~T and obstacles placed at 607~nm, 407~nm, and 207~nm from the injection site. It is visibly apparent that as the obstacle is closer to the injection site, the current density range of the VF and VB regimes widens. We quantitatively determine the transition current densities by a sudden change in the spectral peak amplitude. These current densities are plotted in Fig.~\ref{fig5}(b) as a function of obstacle distance and color-coded by the external field. There is an asymmetry between positive and negative transition current densities leading to VB and VF regimes, respectively. However, in both cases, the transition occurs at lower current densities as the obstacle is closer to the injection site. One must note that these current densities are approximate because of the time delay between vortex shedding and its impact on the dynamics at the spin pumping site. Indeed, vortex motion on a textured magnetization undergoes Kelvin motion coupled with vortex-vortex and vortex-antivortex interactions that include sinuous and varicose modes as well as rotations about a vorticity center~\cite{Iacocca2017,Iacocca2017b}.

We also note that the VF regime was not observed for $\mu_0H_0=0.2$~T. This is because the DEF at negative currents induces a torque that tilts the density $n$, or $m_z$, against the external field. As a result, $n$ is close to zero, c.f. Fig.~\ref{fig3}(b), and the flow is mostly laminar. It is well-established within spin hydrodynamics that the flow is subsonic when $n=0$ up to sub-exchange-length wavelengths when modulational instability~\cite{Iacocca2017} or, equivalently, the Landau criterion~\cite{Sonin2019} ensues.
\begin{figure}[t]
\centering \includegraphics[trim={0in 0in 0.2in 0in}, clip, width=3.4in]{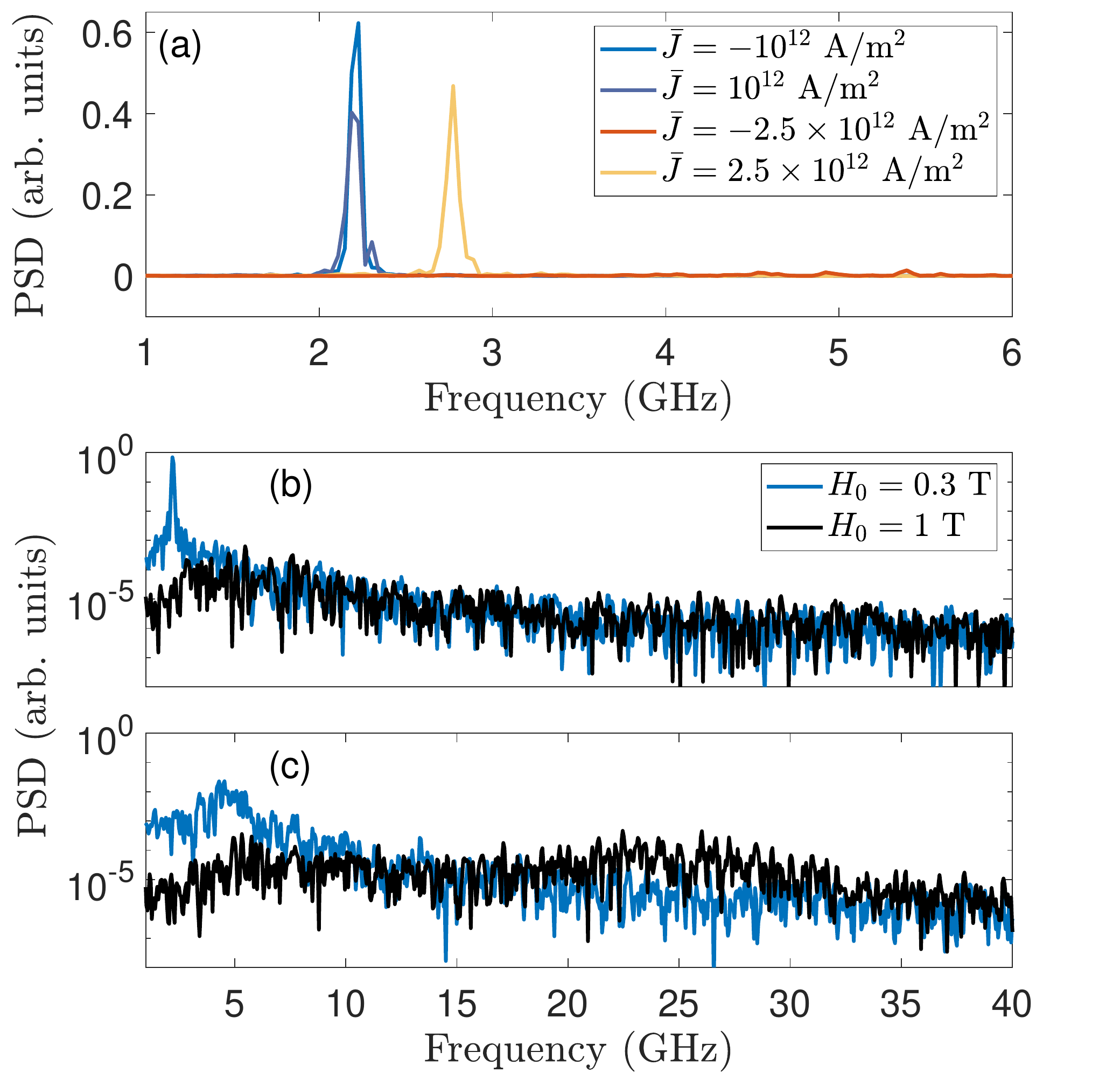}
\caption{ \label{fig5.2} (a) Spectra of the spatially averaged in-plane magnetization at the spin pumping site in simulations including thermal fluctuations at room temperature ($300$~K). An obstacle is placed $307$~nm from the injection site and the external field is set to $\mu_0H_0=0.3$~T. The LF regime is obtained for both $\bar{J}=-10^{12}$~A/m$^2$ and $\bar{J}=10^{12}$~A/m$^2$ (blue and dark blue curves) displaying peaks at $\approx2.2$~GHz. The VF regime is obtained for $\bar{J}=-2.5\times10^{12}$~A/m$^2$ and displays no visible peak in comparison with the spectrum from the LF regime. The VB regime is obtained for $\bar{J}=2.5\times10^{12}$~A/m$^2$ and displays a clear peak at $\approx2.8$~GHz. The spectrum for $\bar{J}=-10^{12}$~A/m$^2$ is shown in (b) by the blue curve in logarithmic scale in amplitude and frequency axis up to $40$~GHz. The spin wave contribution to the spectrum can be computed by setting a saturating field of $\mu_0H_0=1$~T (black curve). In this case, the spin wave contribution is hidden under the noise floor, indicating incoherent magnetization precession at the detector's site. In (c), we set $\bar{J}=-4\times10^{12}$~A/m$^2$. At $\mu_0H_0=0.3$~T (blue curve), a broad peak at $\approx5$~GHz corresponds to the VF regime's detected dynamics. At $\mu_0H_0=1$~T (black curve), the spin wave contribution is now visible as a broad peak at $\approx25$~GHz. }
\end{figure}
\begin{figure*}[t]
\centering \includegraphics[trim={1.2in 0.in .5in 0.in}, clip, width=7in]{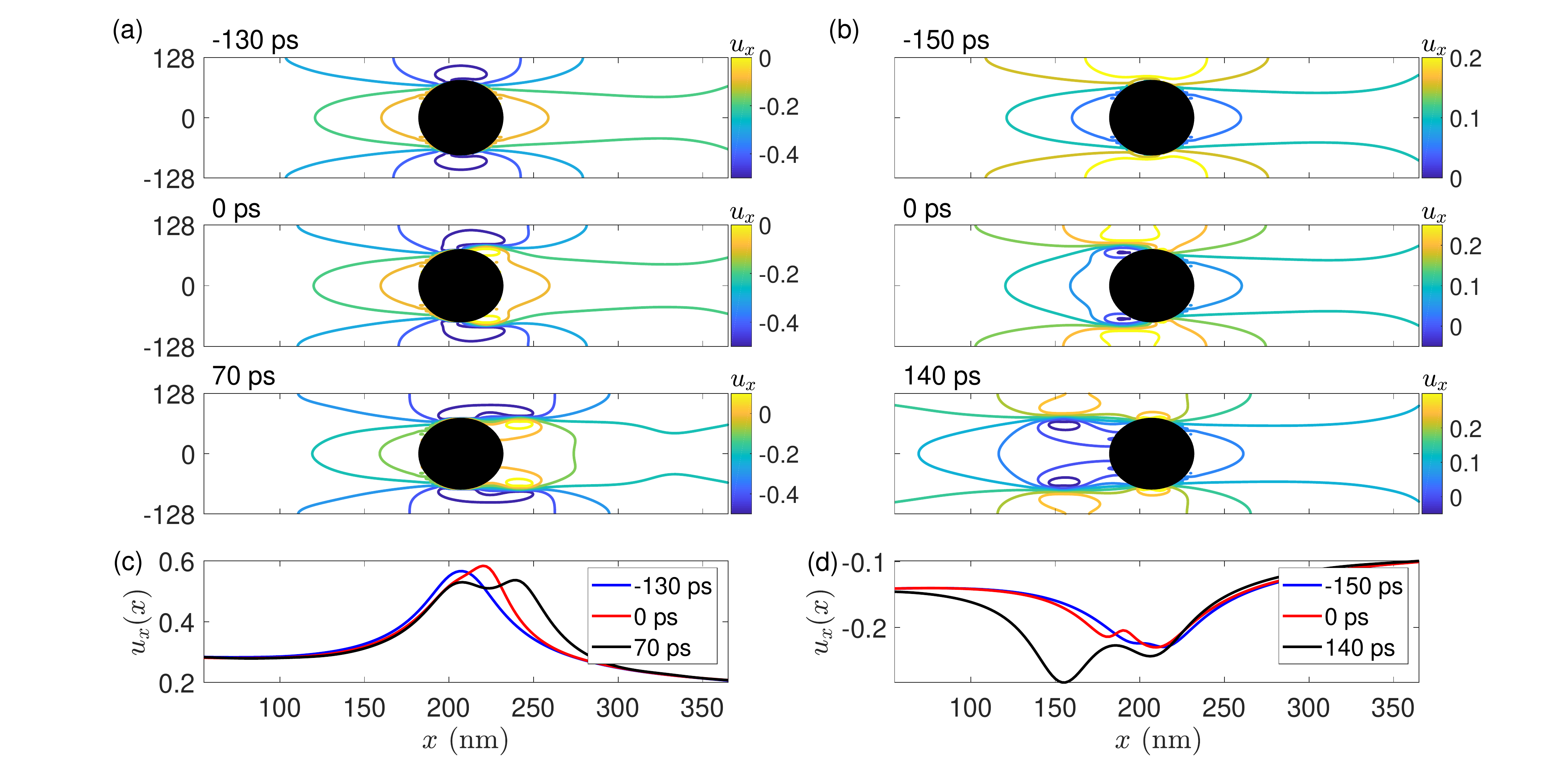}
\caption{ \label{fig6} Vortex shedding dynamics in the (a) VF and (b) VB regimes shown by contours of the $x$ component of the fluid flow, $u_x$. The obstacle is elliptical but appears circular due to the figure's aspect ratio. In (a), the increasing flow magnitude at the edge of the obstacle leads to the development of a back-flow into the obstacle that generates a V-AV pair. In (b) the vortex sheds because of an accelerating flow that also backflows into the obstacle. Both dynamics are detailed in (c) and (d) as cuts of the $x$ component of the fluid flow, $u_x(x,y_0)$ at $y_0=90$ nm. The snapshots are selected in each case for clarity and indicated relative to the instant where vortex shedding is qualitatively observed.}
\end{figure*}

The regimes described above are robust to thermal fluctuations. This implies that thermally excited spin waves and those scattered off the obstacle do not destabilize the DEF nor change its salient features. For example, we show in Fig.~\ref{fig5.2}(a) the spectra of the spatially averaged in-plane magnetization when $\mu_0H_0=0.3$~T, thermal fluctuations at room temperature $(300$~K) are included in the simulations, and the obstacle is placed at $307$~nm from the injection site. The spectra is obtained from $20$~ns time-traces with a resolution of $37$~MHz. For $\bar{J}=10^{12}$~A/m$^2$ and $\bar{J}=-10^{12}$~A/m$^2$, the DEF is in a LF regime, and both spectra (blue and dark blue curves) are essentially identical at $\approx2.2$~GHz, as expected from symmetry. In the VF regime when $\bar{J}=-2.5\times10^{12}$~A/m$^2$ (red curve), the long-range coherence is destroyed and the spectrum is essentially noise. In the VB regime when $\bar{J}=2.5\times10^{12}$~A/m$^2$ (yellow curve), we recover a clear peak at approximately $2.8$~GHz, in agreement with the constant frequency for the same external field shown in Fig.~\ref{fig4}.

The spin waves scattered off the obstacle lose coherence because of wave interference. This is markedly different to DEFs whose coherence is guaranteed in the LF regime and, partially, in the VB regime. As a result, the spectrum from a pure spin-wave contribution at the detector site will exhibit significant spectral broadening. This is observed in Fig.~\ref{fig5.2}(b) and (c) where we perform the simulations at room temperature described above as a function of field. In panel (b), the current is set to $\bar{J}=-2.5\times10^{12}$~A/m$^2$. The blue curve is obtained at $\mu_0H_0=0.3$~T and it corresponds to the blue curve in panel (a) now shown in natural logarithmic scale and a frequency range up to $40$~GHz. When the field is set to $\mu_0H_0=1$~T, the magnetization is saturated in the out-of-plane orientation, and DEFs are suppressed~\cite{Iacocca2019b}. Therefore, the resulting spectrum is only due to current and thermally excited spin waves. Clearly, there is no discernible peak, indicating an incoherent magnetization precession at the detector site. In panel (c), $\bar{J}=-4\times10^{12}$~A/m$^2$ and the DEF is in a VF regime at $\mu_0H_0=0.3$~T. We observe a broad peak at $\approx5$~GHz. Upon saturation, this large-amplitude current density can excite spin waves that do reach the detector site but are nonetheless largely incoherent. This is evidenced by the broad peak located at $\approx25$~GHz. These simulations demonstrate that obstacles can be considered to be efficient spin-wave scatterers and used to distinguish DEFs or spin superfluids from spin waves.

\section{Vortex shedding dynamics}

We study the vortex shedding process at the transition between laminar flow and both VF and VB regimes. Vortex shedding from a uniform magnetic texture interacting with an obstacle was studied in Ref.~\onlinecite{Iacocca2017b}. A phase diagram for vortex shedding and the different V-AV train dynamics was obtained in that work, returning a non-trivial dependence between the obstacle size and the impinging fluid velocity. However, the general process can be qualitatively understood as analogous to fluids~\cite{Williamson1996,Leweke2016} or even superfluids~\cite{Reeves2015,Kwon2016}. A V-AV pair is nucleated when the fluid flow at the edges of the obstacle becomes locally supersonic while the flow behind the obstacle remains subsonic. This causes a back-flow into the obstacle that gives rise to vortices of opposite circulation at each extremum of the obstacle. Such vortices compose a V-AV pair in the magnetization.

The V-AV pair nucleation dynamics from a DEF is shown in Fig.~\ref{fig6} when the applied field is $\mu_0H_0=0.3$~T and the obstacle is placed $207$~nm from the injection site. The series of snapshots show the contour lines of $u_x(x,y)$ in the onset of the VF regime at $\bar{J}=-3.1\times10^{12}$~A/m$^2$ in Fig.~\ref{fig6}(a), and the onset of the VB regime at $\bar{J}=2.2\times10^{12}$~A/m$^2$ in Fig.~\ref{fig6}(b). Note that the axes are not in scale and the obstacle appears circular. In both cases, the V-AV pair is nucleated at the extrema of the obstacle upon reversal of the flow in the obstacle's wake. Each vortex is characterized by circular contours of $u_x(x,y)$.

The asymmetry in the critical current for vortex shedding can be qualitatively understood from these dynamics. In Fig.~\ref{fig6}(c) and (d), we show $u_x(x,y_0)$ with $y_0=90$~nm, just 15~nm off the obstacle's edge. In the VF regime shown in (c), the fluid velocity locally increases at $130$~ps before shedding (blue curve). At the time of shedding, the fluid velocity steepens towards a foldover profile (red curve). Foldover cannot occur because of magnetic exchange so that a vortex is nucleated, visible at $70$~ps after the shedding (black curve). In this process, the fluid flow downstream is lower in magnitude so that a significant flow must accumulate upstream for a vortex to be shed. In the VB regime shown in (d), the evolution is qualitatively different. Here, the fluid flow is larger in magnitude upstream. As a consequence, the flow does not accumulate to the point of steepening but instead sheds a vortex as the fluid flow is accelerated past the obstacle.

The vortex shedding dynamics described above can occur for obstacles of different dimensions. As demonstrated in previous works~\cite{Sonin2010,Takei2014,Iacocca2019b}, the most relevant length scale for an ideal DEF or spin superfluid is the nanowire's length, where the precessional frequency is inversely proportional to it. However, when including an obstacle, the nanowire's width and the obstacle's dimensions relative to the fluid flow also become important lengthscales for vortex shedding. To explore such a dependency, we perform simulations with elliptical obstacles with major and minor axes dimensions ranging from $10$~nm to $150$~nm in steps of $20$~nm. To reduce the parameter space, we choose an applied field of $\mu_0H_0=0.5$~T and we consider two input currents, $\bar{J}_1=-3\times10^{12}$~A/m$^2$ and $\bar{J}_2=-4\times10^{12}$~A/m$^2$. These simulations allow for the determination of the VF transition as a function of the obstacle's aspect ratio.

Based on the vortex shedding dynamics described in Fig.~\ref{fig6}, one expects fluid accumulation at the edges of the obstacle to be the defining parameter to induce local supersonic flow. For an elliptical obstacle, this implies that flow accumulation should be proportional to the curvature of the obstacle's surface with respect to the impinging flow. In other words, an obstacle elongated parallel to the flow will only slightly perturb such flow while an obstacle elongated perpendicular to it will strongly divert the impinging flow. Because the curvature of an ellipse is $\mathcal{C}=a_y/a_x^2$, where $a_y$ and $a_x$ are the ellipse's axes in the $y$ and $x$ direction, the critical flow $u_c$ for the transition as a function of the ellipse's axis and input flow $\bar{u}$ can be assumed to be related by
\begin{equation}
\label{eq:crit_ellipse}
   a_y = \frac{u_c}{\bar{u}}a_x^2+a_\mathrm{min},
\end{equation}
where $a_\mathrm{min}$ is a constant that, in general, depends on the input flow and represents vortex shedding from a one-dimensional barrier.

\begin{figure}[t]
\centering \includegraphics[trim={0in 0.in 0.2in 0.2in}, clip, width=3.4in]{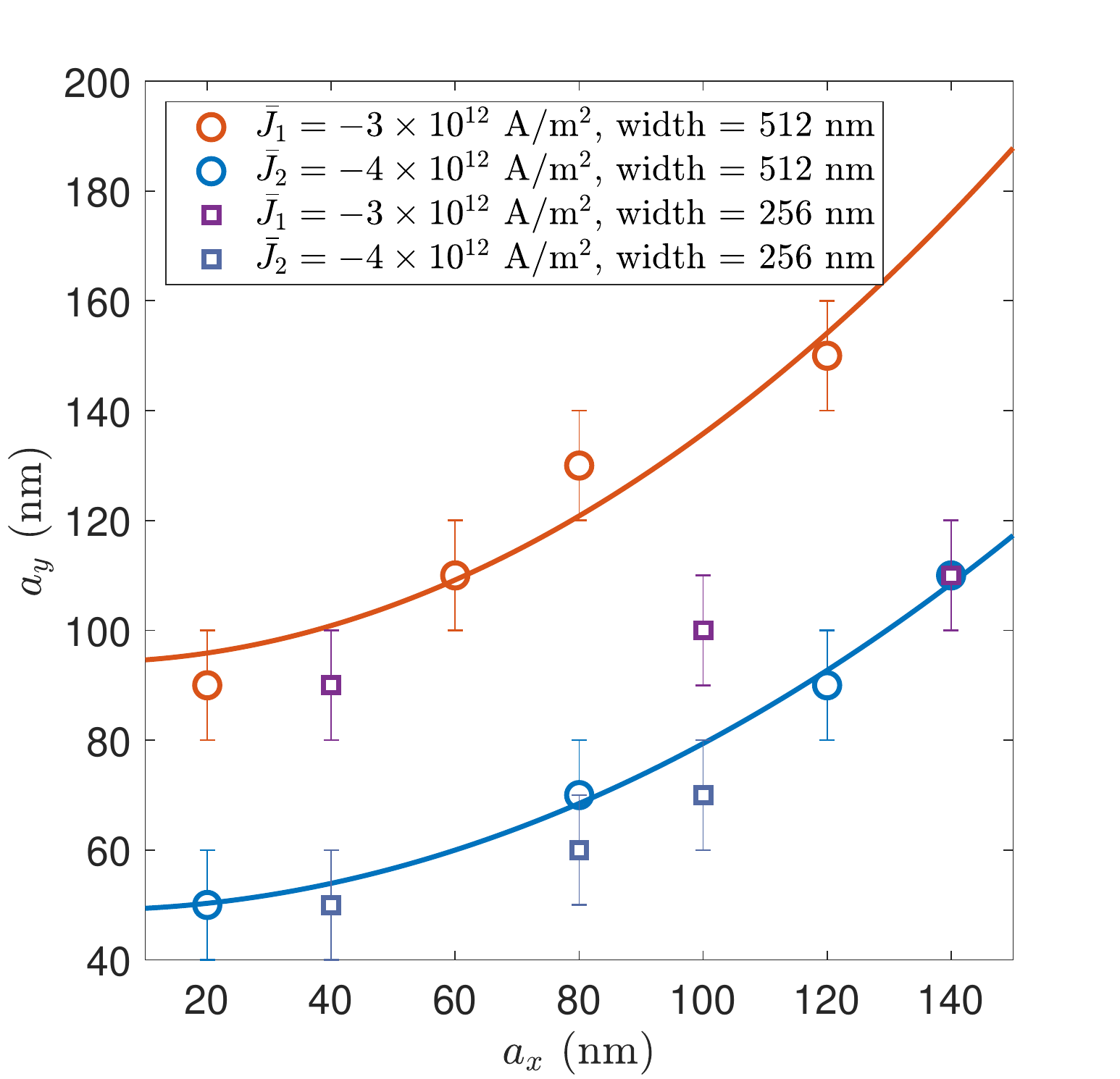}
\caption{ \label{fig9} Vortex shedding transition as a function of the elliptical obstacle's axes, $a_x$ and $a_y$, centered at a position such that the distance to the edge closest to the injection site was fixed at $200$~nm. For a nanowire of dimensions $1024$~nm$\times512$~nm$\times5$~nm, the transitions for currents $\bar{J_1}=-3\times10^{12}$~A/m$^2$ and $\bar{J_2}=-4\times10^{12}$~A/m$^2$ are shown by red and blue circles, respectively. The parabolic fits according to Eq.~\eqref{eq:crit_ellipse} are shwon with the same color code. For a nanowire of dimensions $1024$~nm$\times256$~nm$\times5$~nm, the transitions for currents $\bar{J_1}=-3\times10^{12}$~A/m$^2$ and $\bar{J_2}=-4\times10^{12}$~A/m$^2$ are shown by magenta and indigo squares, respectively. These results suggest a saturation effect induced by the proximity of the nanowire's physical edge. }
\end{figure}
The numerically estimated transitions as a function of the ellipse's axes for a nanowire of dimensions $1024$~nm$\times512$~nm$\times5$~nm and input current densities $\bar{J_1}=-3\times10^{12}$~A/m$^2$ and $\bar{J_2}=-4\times10^{12}$~A/m$^2$ are shown in Fig.~\ref{fig9} by red and blue circles, respectively. In these simulations, the obstacles were positioned such that the edge closest to the injection site was fixed at a distance of $200$~nm. The estimated transition can be well fit with Eq.~\eqref{eq:crit_ellipse}. In addition, it is visibly apparent that the parabola is steeper for the lower current density. Taking the ratio of the concavity of both parabola from Eq.~\eqref{eq:crit_ellipse} and assuming that $\bar{u}\propto\bar{J}$~\cite{Iacocca2019b}, we find a ratio $(u_c/\bar{u}_1)/(u_c/\bar{u}_2)=\bar{J}_2/\bar{J}_1=4/3\approx1.33$. From the fits, we obtain a ratio of $1.37\pm0.3$, in excellent agreement with the analytical estimate. The length of a one-dimensional barrier for vortex shedding, $a_\mathrm{min}$, is fit to $94$~nm~$\pm6$~nm for $\bar{J}_1$ and $49$~nm~$\pm2$~nm for $\bar{J}_2$. These results qualitatively agree with the notion that the length of the barrier is proportional to the flow's wavelength to induce vortex shedding, c.f. wavelenghts of $\approx147$~nm for $\bar{J}_1$ and $\approx105$~nm for $\bar{J}_2$.
\begin{figure}[b]
\centering \includegraphics[trim={0.15in 0in .5in 0.2in}, clip, width=3.3in]{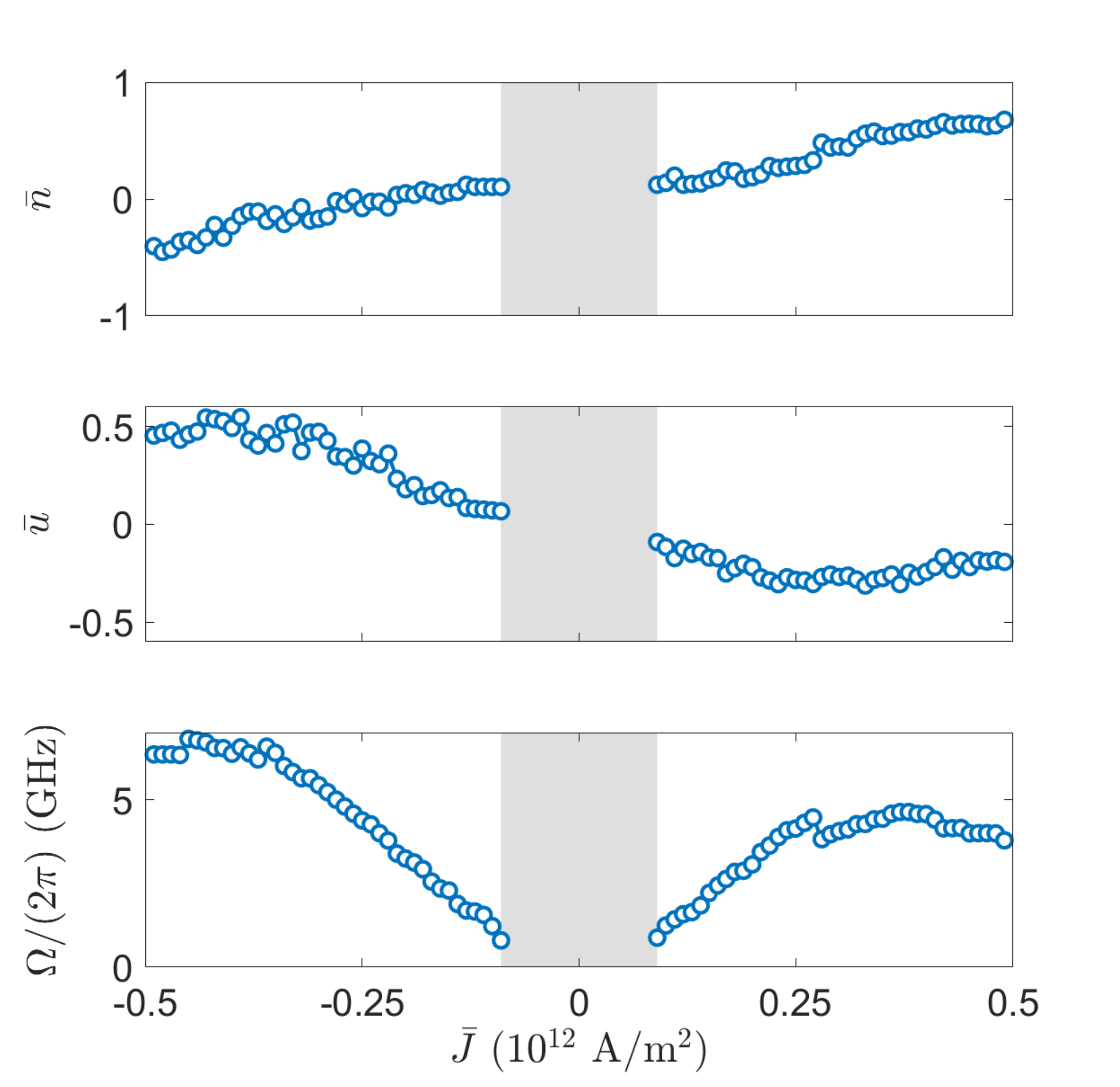}
\caption{ \label{fig7} Current-dependent DEF parameters with an external field of $\mu_0H_0=0.1$~T in a simulated nanowire of thickness 1~nm and considering cubic anisotropy. The sub-threshold region is identified by a gray area. }
\end{figure}

The transition estimated from a nanowire of dimensions $1024$~nm$\times256$~nm$\times5$~nm and input current densities $\bar{J_1}$ and $\bar{J_2}$ are shown in Fig.~\ref{fig9} by magenta and indigo squares, respectively. The indigo squares approximately follow the parabolic fit up to $a_y=70$~nm. However, for $a_y=80$~nm, vortex shedding was observed even for $a_x=150$~nm. This is an indication of a saturation effect induced by the proximity of the nanowire's physical boundary. Such saturation is more clearly observed from the magenta squares that significantly deviate from the transition of the wider nanowire. A detailed analytical formulation for saturation effects and the minimal barrier length is beyond the scope of this work. However, the presented simulations provide a qualitative understanding of these effects.

\section{Vortex shedding in the presence of anisotropy and non-local dipole field}
\begin{figure}[b]
\centering \includegraphics[trim={.2in 0in .2in 0.2in}, clip, width=3.4in]{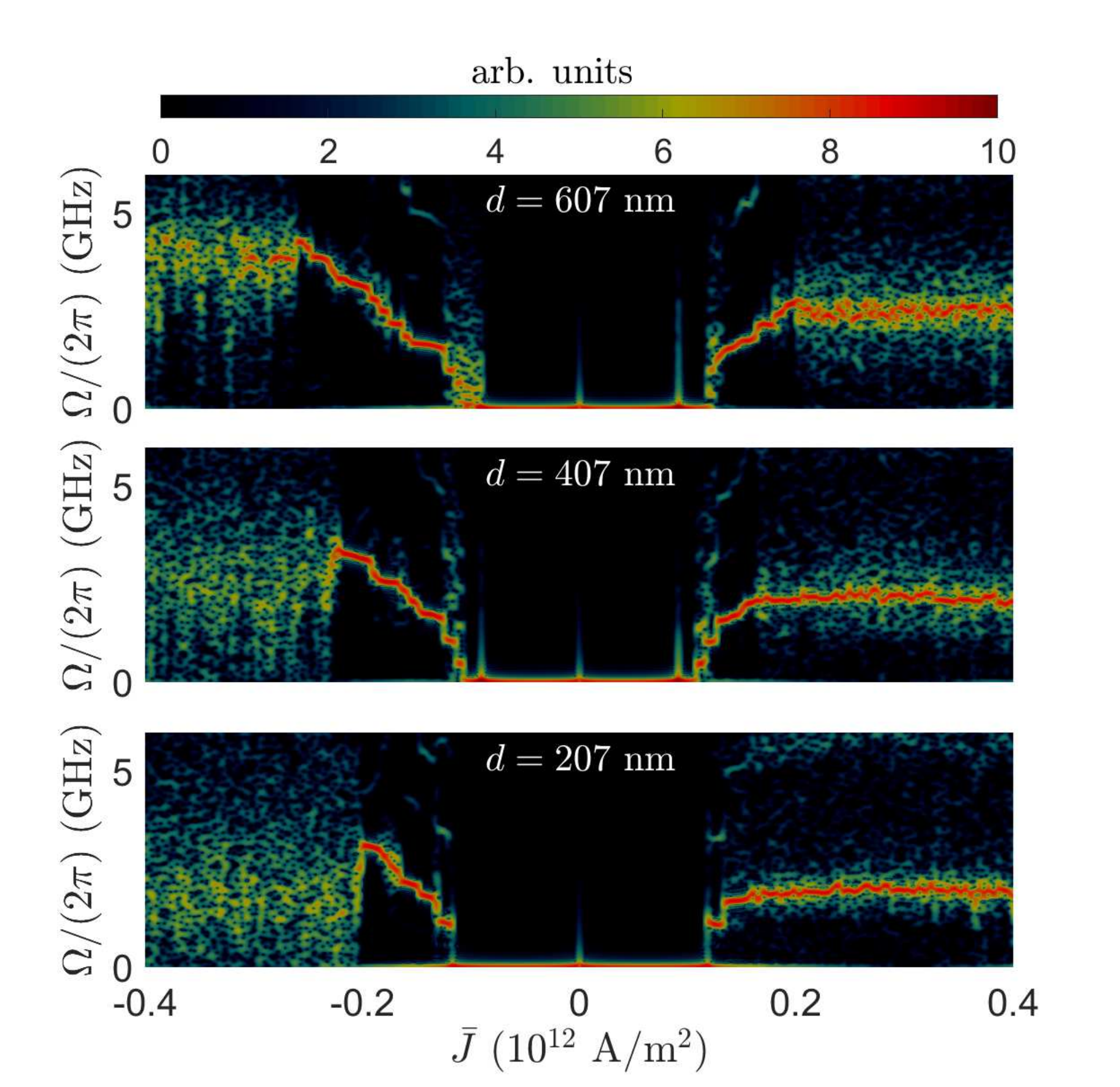}
\caption{ \label{fig8} DEF frequency as a function of current when $\mu_0H_0=0.1$~T in a 1~nm-thick nanowire with anisotropy. The obstacle is placed at $607$~nm, $407$~nm, and $207$~nm from the injection site. The VF and VB regimes' range increases as the obstacle is placed closer to the injection site, similar to the ideal case.}
\end{figure}

We include the contributions of the non-local dipole and anisotropy fields in the micromagnetic simulations. The non-local dipole field depends on the thickness and may drive the DEF unstable. The study of such instabilities is beyond the scope of this work. Therefore, we restrict ourselves to simulating a nanowire of 1~nm in thickness so that dipole fields are minimized. The anisotropy of permalloy is modeled by grains with randomly oriented first-order cubic anisotropy and anisotropy constant $K_c=$3,950~J/m$^3$. We consider grains with an average size on the order of the exchange length. The spin injection and spin pumping sites are extended to 20~nm in these simulations. We also restrict the simulations to the case of an external field of $\mu_0H_0=0.1$~T.

The DEF parameters as a function of current are shown in Fig.~\ref{fig7}. The qualitative features are similar to those of the ideal nanowire shown in Fig.~\ref{fig3}. However, the current densities are significantly lower. This is because of the increased spin injection area compared to the thickness of the film~\cite{Iacocca2019b}. The anisotropy also imposes a current density threshold, marked as a gray area. We note that a linear regime is observed between $\bar{J}\approx-0.35\times10^{12}$~A/m$^2$ and $\bar{J}\approx0.25\times10^{12}$~A/m$^2$.

The DEF's frequency as a function of current is obtained from a SPWV distribution analysis of simulations where the current is swept between $0$ to either $-0.5\times10^{12}$~A/m$^2$ or $0.5\times10^{12}$~A/m$^2$. The simulations run for 500~ns. The resulting SPWV distribution for obstacles placed at $607$~nm, $407$~nm, and $207$~nm from the injection site are shown in Fig.~\ref{fig8}. The previously identified VF, LF, and VB regimes are observed.

These simulations demonstrate that the features studied in sections III and IV are maintained in the presence of anisotropy and weak non-local dipole field.

\section{Conclusion}

We studied the characteristics of DEFs excited by spin injection in magnetic nanowires with a physical obstacle. The fluid-like properties of DEFs allow for a regime of laminar flow around the obstacle that maintains their long-range coherence. This is in contrast to spin waves that scatter off a physical boundary. Therefore, obstacles can be thought of as control gates that favor long-range DEFs over spin-wave propagation in nanowires. The marked differences in coherence would allow to experimentally disentangle whether the signal detected by, e.g., inverse spin-Hall effect originates from either DEFs or spin waves.

In addition, numerical simulations demonstrated vortex shedding as a function of spin injection, leading either to a VF regime where V-AV pairs translate to the spin pumping site and break the long-range coherence of DEFs, or a VB regime where V-AV pairs translate back to the injection site and favor a specific DEF frequency at the spin pumping site. These two regimes offer qualitative features to recognize experimentally.

The results presented here are not limited to ferromagnetic materials. Previous theoretical~\cite{Takei2014,Takei2014b,Sonin2019} and numerical~\cite{Evers2020} studies have shown that the spin superfluid equations are obtained for both ferromagnets and antiferromagnets and experiments seeking spin superfluidity have so far focused on antiferromagnetic materials~\cite{Yuan2018,Stepanov2018}. Antiferromagnets atomically compensate non-local dipole fields, effectively eliminating any instability related to the sample's thickness and saturation magnetization. It would be interesting to study vortex shedding~\cite{Sonin2019} and their dynamics in antiferromagnetic nanowires to establish their impact on the long-range coherence of spin superfluids, and more generally, on DEFs. In addition, the use of obstacles would also aid to distinguish detected signals from thermally excited spin waves away from the injection site~\cite{Lebrun2018} and the signals produced by DEFs.

Ferrimagnetic materials would be also interesting to study because their reduced saturation magnetization minimizes the impact of non-local dipole fields. Recent experiments on long-distance transport in YIG and disordered a-YIG~\cite{Wesenberg2017} are promising to seek spin superfluid and DEF solutions in amorphous magnets~\cite{Ochoa2018}. The use of obstacles could also help in this case to distinguish the origin of long-range spin transport.

In summary, our results demonstrate that obstacles can be used to control vortex shedding from DEFs in nanowires. This basic form of control can serve as the basis for more complex manipulation of DEFs and, therefore, of the transport of angular momentum through magnetic materials.

\begin{acknowledgments}
E.I. is grateful to Mark Hoefer for providing access to the University of Colorado Blanca cluster and acknowledges support from the U.S. Department of Energy, Office of Science, Office of Basic Energy Sciences under Award Number DE-SC0017643.
\end{acknowledgments}


%

\end{document}